\documentclass[%
 reprint,
showpacs,
 amsmath,amssymb,
aps,
pre,
]{revtex4-1}

\usepackage{amsfonts}
\expandafter\let\csname equation*\endcsname\relax
 \expandafter\let\csname endequation*\endcsname\relax
\usepackage{amsmath}
\usepackage{amssymb}
\usepackage{graphicx}
\usepackage{bm}
\usepackage{hyperref}
\usepackage[usenames]{color}
\usepackage{subfigure}
\def\dbar{{\mathchar'26\mkern-12mu d}} 
\begin{document}
\title{Rabi model as a quantum coherent heat engine: from quantum biology to superconducting circuits}
\author{Ferdi Altintas$^1$}
\author{Ali \"U. C. Hardal$^2$} 
\author{\"{O}zg\"{u}r E. M\"{u}stecapl{\i}o\u{g}lu$^2$}
\email{omustecap@ku.edu.tr}
\affiliation{$^1$ Department of Physics, Abant Izzet Baysal University, Bolu, 14280, Turkey}
\affiliation{$^2$ Department of Physics, Ko\c{c} University, Sar{\i}yer, \.Istanbul, 34450, Turkey}
\begin{abstract}
We propose a multilevel quantum heat engine with a working medium described by a generalized Rabi model which consists of a two-level system coupled to a single mode bosonic field. The model is constructed to be a continuum limit of a quantum biological description of light harvesting complexes so that it can amplify quantum coherence by a mechanism which is a quantum analog of classical Huygen's clocks.
The engine operates in quantum Otto cycle where the working medium is coupled to classical heat baths in the isochoric processes 
of the four stroke cycle; while either the coupling strength or the resonance frequency is changed in the adiabatic stages.  
We found that such an engine can produce work with an efficiency close to 
Carnot bound when it operates at low temperatures and in the ultrastrong coupling regime. Interplay of quantum coherence and quantum correlations on the engine performance is discussed in terms of second order coherence, quantum mutual information and logarithmic negativity of entanglement.
We point out that the proposed quantum Otto engine can be implemented experimentally with the modern circuit 
quantum electrodynamic systems where flux qubits can be coupled ultrastrongly to superconducting transmission line resonators.  
\end{abstract}
\maketitle
\section{Introduction}
\label{intro}
Quantum heat engines (QHEs) have attracted much attention recently~\cite{maruyama_colloquium:_2009,scully_quantum_2010,
huang_effects_2012,tonner_autonomous_2005,wang_thermal_2009,goswami_thermodynamics_2013,kieu_second_2004,
quan_quantum_2007,allahverdyan_work_2005,wang_quantum_2012}, due to growing need for energy harvesting, storage and transfer efficiently with quantum devices. They raise fundamental questions about generalizing thermodynamics to quantum regime and to quantum information. Some quantum systems, such as masers, have already been identified as quantum heat engines, which can operate at Carnot efficiency, long time ago ~\cite{scovil_three-level_1959}. Though they are not realized yet, QHEs are proposed for ultracold atoms~\cite{fialko_isolated_2012}, optomechanics~\cite{zhang_quantum_2014,zhang_theory_2014}, quantum dots~\cite{bergenfeldt_hybrid_2014}, quantum Hall edge states~\cite{sothmann_quantum_2014}, and superconducting circuits~\cite{quan_maxwells_2006}. Modern ideas revealed that Carnot efficiency can be beaten in QHEs, without violating the second law of thermodynamics, by exploiting quantum coherence~\cite{scully_quantum_2010} or by quantum reservoir engineering~\cite{huang_effects_2012,rosnagel_nanoscale_2014}. Such proposed QHEs require highly sophisticated quantum engineering techniques and their dependence on quantum coherence makes them susceptible to quantum decoherence~\cite{quan_quantum-classical_2006}. 

Nature, on the other hand, has its subtle ways to utilize quantum coherence even at physiological temperatures in energy processes essential for life, such as photosynthesis. Recent studies revealed that long range quantum coherence~\cite{sumi_bacterial_2001,engel_evidence_2007,collini_coherently_2010} and associated quantum correlations~\cite{caruso_entanglement_2010,sarovar_quantum_2010,ishizaki_quantum_2010,fassioli_distribution_2010} play crucial role in light harvesting complexes~\cite{cogdell_architecture_2006,hu_photosynthetic_2002}. Such systems can be described as dipole interacting spins~\cite{cheng_dynamics_2009,renger_ultrafast_2001}. Motivated by the natural solution to interaction enhanced quantum coherence, we propose a biomimetic design~\cite{sarovar_design_2013} of an interacting effective spin system as a working medium for a synthetic 
quantum coherent heat engine. 

Establishing long range quantum coherence by homogeneously coupling many spins directly to each other is challenging in synthetic systems. We propose an alternative indirect path to coherence, inspired by the Huygen’s clocks which can be synchronised when they are suspended from a common beam~\cite{bennett_huygenss_2002}. The imperceptible movements of the beam, acting like a conductor of an orchestra, build coherence between the clocks. We envision a scheme where large number of spins, acting like quantum Huygen’s clocks, are homogeneously coupled to a central spin, which could build a global quantum coherence. 

We examine an extreme but naturally occuring case in light harvesting complexes where only few spins in the large ensemble can be excited~\cite{fassioli_energy_2009}. Such collective excitations can be described by a bosonic field of spin waves~\cite{holstein_field_1940,dyson_general_1956}. The interaction of the spin ensemble with the central spin is then collectively enhanced and mutated into a spin-boson form known as Rabi model~\cite{rabi_process_1936,rabi_space_1937,braak_integrability_2011}.  Implementing the natural strategy of light harvesting complexes indirectly with a continuum of quantum Huygen’s clocks, Rabi model synthetic working substance could build interaction enhanced global coherence for optimum and efficient work harvesting. 
We assume the working medium of the engine undergoes a quantum Otto cycle~\cite{kosloff_discrete_2002,kieu_second_2004,quan_quantum_2007}. 
Quantum Otto engines considered in the literature
with non-interacting~\cite{abah_single-ion_2012,rezek_irreversible_2006} or interacting working medium~\cite{zhang_entangled_2008,wang_thermal_2009,thomas_coupled_2011,huang_special_2014,PhysRevE.90.032102,zhang_four_2007} are limited to few level coupled spin systems. A biomolecular Carnot cycle has been proposed to explain operation of photosynthetic reaction centers recently~\cite{dorfman_photosynthetic_2013}.

Our analysis revealed that global coherence in the system can be built and enhanced with the spin-boson coupling. We find that there is a critical interaction at which the interplay of coherence and quantum correlations yields maximum work and efficiency. Requirement of a critical coherence for optimum work has also been reported for a single four-level atom with noise induced quantum coherence (Fano effect) between its two levels~\cite{goswami_thermodynamics_2013,huang_special_2014}. Our case generalizes this effect to interaction enhanced multi-level coherence. The critical amount of coherence is found to be available in the ultra strong interaction regime, where coupling is comparable to excitation energy. This regime complies with the Rabi model description and experimentally available in superconducting circuit quantum electrodynamics (QED) systems~\cite{forn-diaz_observation_2010,niemczyk_circuit_2010}. We propose such a system for a compact implementation of our model and estimate that its power output could be significantly greater than non-interacting systems.
\section{Construction of physical model of quantum coherent working substance}
\label{model}
\begin{figure}[!t]
\begin{center}
\includegraphics[width=8.0cm]{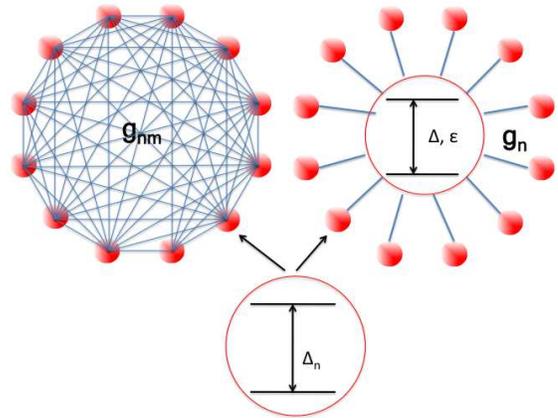}
\caption{\label{fig:lhc} (Color online) Schematic structures of light harvesting antenna complexes of $N$ pigments (choromophores). Each pigment is a two level system with an energy separation of $\Delta_n$. (Left complex) Pigments are interacting with each other with a coupling coefficient of $g_{mn}$. (Right complex) Non-interacting pigments are coupled to a central one with an interaction coefficient $g_n$. The central pigment carries a small quantum coherence, characterized by $\epsilon$.}
\end{center}
\end{figure}

We apply a biomimetic strategy to construct our physical model of quantum coherent working substance~\cite{sarovar_design_2013}. 
Our inspiration comes from
the model of $N$ interacting chromophores (pigments) in light harvesting photosynthetic complexes, whose electronic Heitler-London Hamiltonian, $H_e$, is described by ($\hbar=1$)~\cite{cheng_dynamics_2009,renger_ultrafast_2001}
\begin{eqnarray}
H_e=\sum_{n=1}^{N}\Delta_n\sigma_n^\dag\sigma_n+\sum_{n>m}^{N}g_{nm}(\sigma_n^\dag\sigma_m+\sigma_m^\dag\sigma_n),
\end{eqnarray}
where $\Delta_n$ is the excitation energy of the localized two-level chromophore at site $n$ and $g_{nm}$ is the dipolar coupling describing the electronic excitation transfer between chromophores at sites $n$ and $m$. The Pauli spin operators $\sigma_n$ and $\sigma_n^\dag$ are the annihilation and creation operators of electronic excitations at site $n$. In the complex the chromophores are typically distributed on a ring and all the sites are coupled to each other as depicted in Fig.~\ref{fig:lhc}. 

Two-level pigments distributed on the ring act as an antenna complex and the long range interactions in the system contribute for quantum coherence enhanced solar energy harvesting. Single excitations, Frenkel excitons, are delocalized over the sites for coherent transfer of the energy through the complex. We propose an alternative structure which can establish coherent coupling of two level systems indirectly. Let us consider a single central pigment with a small amount of coherence, $\epsilon$, coupled to a large ensemble of non-interacting pigments. The model Hamiltonian can be described as
\begin{eqnarray}
\nonumber H&=&\Delta\sigma^\dag\sigma+\epsilon(\sigma^\dag+\sigma)+\sum_{n=1}^{N}\Delta_n\sigma_n^\dag\sigma_n\\
&+&\sum_{n=1}^{N}g_{n}(\sigma_n^\dag\sigma+\sigma^\dag\sigma_n),
\end{eqnarray}
where $\Delta$ is the energy of the central site and $g_n$ is the coupling coefficient of the site $n$ to the central one. 
Second term yields a quantum superposition of two levels of the central pigment and establishes a small amount of coherence in the central site. This local coherence is distributed over the other sites by interactions. Participation of many sites to such a delocalized coherence is then amplifies the global coherence of the complex to enhance its operational efficiency. This strategy can be easier to implement in synthetic energy harvesting systems as it would not require long range coupling of all sites.
In light harvesting systems such a coherence can be induced spontaneously by noise as explained by a four level model~\cite{goswami_thermodynamics_2013,huang_special_2014,dorfman_photosynthetic_2013}. Connection of large antenna complexes to relatively smaller reaction centers~\cite{lambert_quantum_2013} or coherent coupling of vibrational bath to excitons~\cite{chin_role_2013} are also 
closely related to our proposed model. 

Distribution of surrounding pigments about the central one can be assumed radially uniform so that 
we can take identical coupling coefficients as $g_n=g_0$. Let us define operators $a$ and $a^\dag$ to describe
annihilation and creation of collective excitations as
\begin{eqnarray}
a=\frac{1}{\sqrt{N}}\sum_{n=1}^N\sigma_n.
\end{eqnarray}
If the number of pigments is much larger than the excitation number, which means $N\gg 1$ for single excitation, 
the operators describe bosonic excitations, analogs of magnons of spin waves~\cite{holstein_field_1940}. This extreme delocalization over the
all sites can be achieved in synthetic systems. The coupling coefficient in this case is collectively enhanced and becomes
$g=\sqrt{N}g_0$. For convenience we take $\Delta_n/2=\omega$ and $\Delta/2\rightarrow \Delta$. Using $\sigma_x=\sigma^\dag+\sigma$ and $\sigma_z=2\sigma^\dag\sigma-1$, the model then becomes
\begin{eqnarray}
H=\Delta\sigma_z+\epsilon\sigma_x+\omega a^\dag a+g(a^\dag\sigma+\sigma^\dag a),
\end{eqnarray}
which is a generalized Jaynes-Cummings model (JCM)~\cite{jaynes_comparison_1963}. 

If the interaction coefficient $g$ can be comparable to excitation energy $\omega$, the JCM should be generalized to a more generic spin-boson model,
known as Rabi model. In our case we then consider a generalized Rabi Hamiltonian which can be written as~\cite{braak_integrability_2011}
\begin{eqnarray}\label{hammilton}
H=\Delta\sigma_z+\epsilon\sigma_x+\omega a^{\dagger}a+g\sigma_x(a+a^{\dagger}),
\end{eqnarray}
The term, $\epsilon\sigma_x$ explicitly breaks the $Z_2$ symmetry of the model when $\epsilon\neq n\omega/2$ ($n$ is an integer)~\cite{braak_integrability_2011}. We take a small value for $\epsilon$ and exploit the avoided level crossings due to this term in arrangement of the numerically determined eigenstates and eigenvalues  $E_n$ of the Hamiltonian~(\ref{hammilton}) in the order of increasing energy, $E_{n+1}>E_n$. We take $\Delta=\omega/2$ and characterize the system by two parameters $(\omega,g)$.

An intuitive appreciation of how this model can establish quantum coherence can be found in a classical analog. Huygen’s clocks, or non-interacting simple pendulums suspended from a common beam, are found to be synchronised. This classical phase synchronization effect is known as the Huygen's odd kind of sympathy~\cite{bennett_huygenss_2002}. The imperceptible movements of the beam, acting like a conductor of an orchestra, build phase coherence between the clocks. Our envisioned scheme could build a global quantum coherence as it consists of large number of two level systems, acting like quantum Huygen’s clocks, which are uniformly coupled to a central coherent oscillator. 
In the subsequent discussions, we consider such a quantum coherent system described by the generalized Rabi model, as a working medium, and examine its work output in a quantum Otto cycle from the perspective of quantum coherence.
\section{Quantum Otto cycle}
\label{sec:cycle}
The quantum Otto cycle consists of four stages. 
The first stage is a quantum isochoric process. The working medium with  $(\omega=\omega_h,g=g_h)$ is brought in contact with a hot thermal reservoir at temperature $T=T_h$. An amount of heat $Q_1$ is absorbed from the heat bath; but no work is done during the stage. The energy levels remain the same $E_n=E_n^h$ while the occupation probabilities change to $P_n(T_h)$ in thermal equilibrium. The second stage is a quantum adiabatic expansion process. The working medium is isolated from the heat reservoir and its energy levels are shifted adiabatically from $E_n^h$ to $E_n=E_n^l$ either by varying $\omega_h$ to $\omega_l$ or by varying  $g_h$ to $g_l$. No heat is transferred; but an amount of work is done during the stage. The occupation probabilities remain unchanged by the adiabatic theorem. The third stage is another quantum isochoric process where the working medium with $(\omega=\omega_l,g=g_l)$  is brought to contact with a cold entropy sink at $T=T_l$. An amount of heat $Q_2$ is released to the reservoir; but no work is done. Attaining thermal equilibrium changes the occupation probabilities to $P_n(T_l)$, but maintains the energy structure with $E_n=E_n^l$.  In the fourth stage, the working medium goes through a quantum adiabatic contraction process. The energy levels are shifted back to $E_n^h$  either by changing $\omega_l$ to $\omega_h$ or $g_l$ to $g_h$. The occupation probabilities are preserved by the quantum adiabatic theorem while work is done due to changing energy levels during the stage. 

The heat transferred and the work performed in the quantum regime can be calculated by the formalism developed in Ref.~\cite{kieu_second_2004}. The first law of thermodynamics for a quantum system with $n$ discrete energy levels can be written as $dU=\dbar Q+\dbar W=\sum_n\{E_ndP_n+P_ndE_n\}$. In this interpretation, an infinitesimal heat transfer corresponds to an infinitesimal change in occupation probabilities, while an infinitesimal work can be performed by an infinitesimal change in the energy levels. The net work done per cycle in the QHE can be determined by $W=Q_1+Q_2$ where
\begin{eqnarray}\label{hwe}
Q_1&=&\sum_nE_n^h(P_n(T_h)-P_n(T_l)),\\
Q_2&=&\sum_nE_n^l(P_n(T_l)-P_n(T_h)).
\end{eqnarray}
Positive work condition $W>0$ indicates the work performed by the QHE per cycle with the efficiency $\eta=W/Q_1$ and $Q_1>-Q_2>0$ is assumed by the second law of thermodynamics. 

We determine the occupation probabilities by Boltzmann distribution $P_n(T)=\exp{(-\beta E_n)}/Z$ where $Z=\sum_n\exp{(-\beta E_n)}$ is the partition function and $\beta=1/k T$ is the inverse temperature  of the heat bath with $k$ being the Boltzmann constant. In order to verify the thermalization assumption, we solve the Bloch-Redfield master equation in Born-Markov approximations 
$\dot\rho=-i [H,\rho]+\mathcal{L}\rho$, which treats the system as a whole in describing its coupling to the environment~\cite{beaudoin_dissipation_2011,ridolfo_nonclassical_2013}, and find the steady state density matrix. The Liouvillian $\mathcal{L}$, which is given in Refs.~\cite{beaudoin_dissipation_2011,ridolfo_nonclassical_2013}, describes the 
modified decoherence and decay rates for the dressed two level system and the bosonic field, as well as the 
coupling to a thermal reservoir. Comparison with the thermal density matrix $\rho_{th}=\exp{(-\beta H)}/Z$  constructed directly by numerical diagonalization of the Hamiltonian in Eq.~(\ref{hammilton}) assures that independent of the choice of rates in $\mathcal{L}$
the steady state solution produces $\rho_{th}$.
\begin{figure}[t!]
     \begin{center}
        \subfigure[]{%
            \includegraphics[width=7cm]{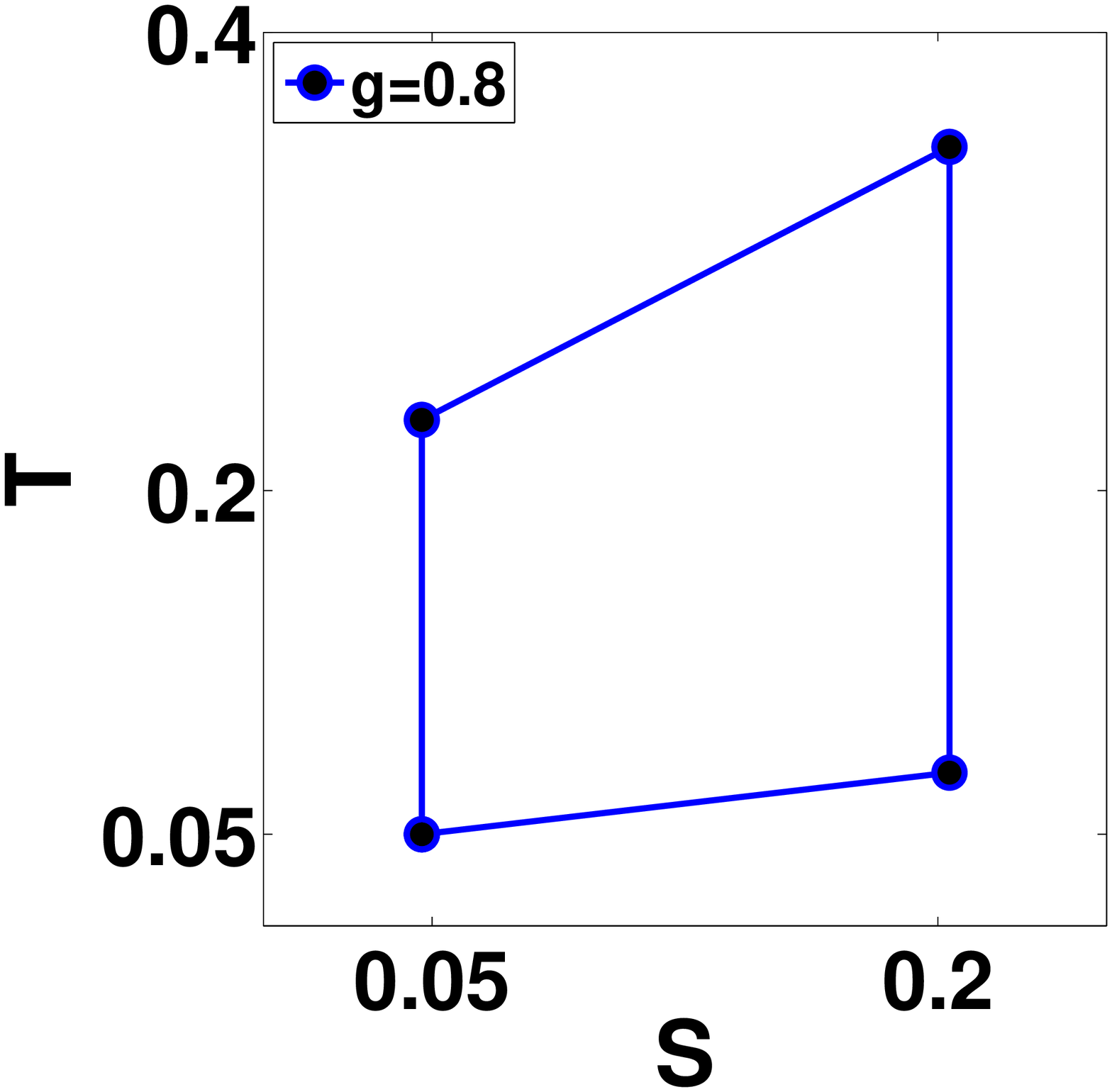}
        }\\%
        \subfigure[]{%
           \includegraphics[width=7cm]{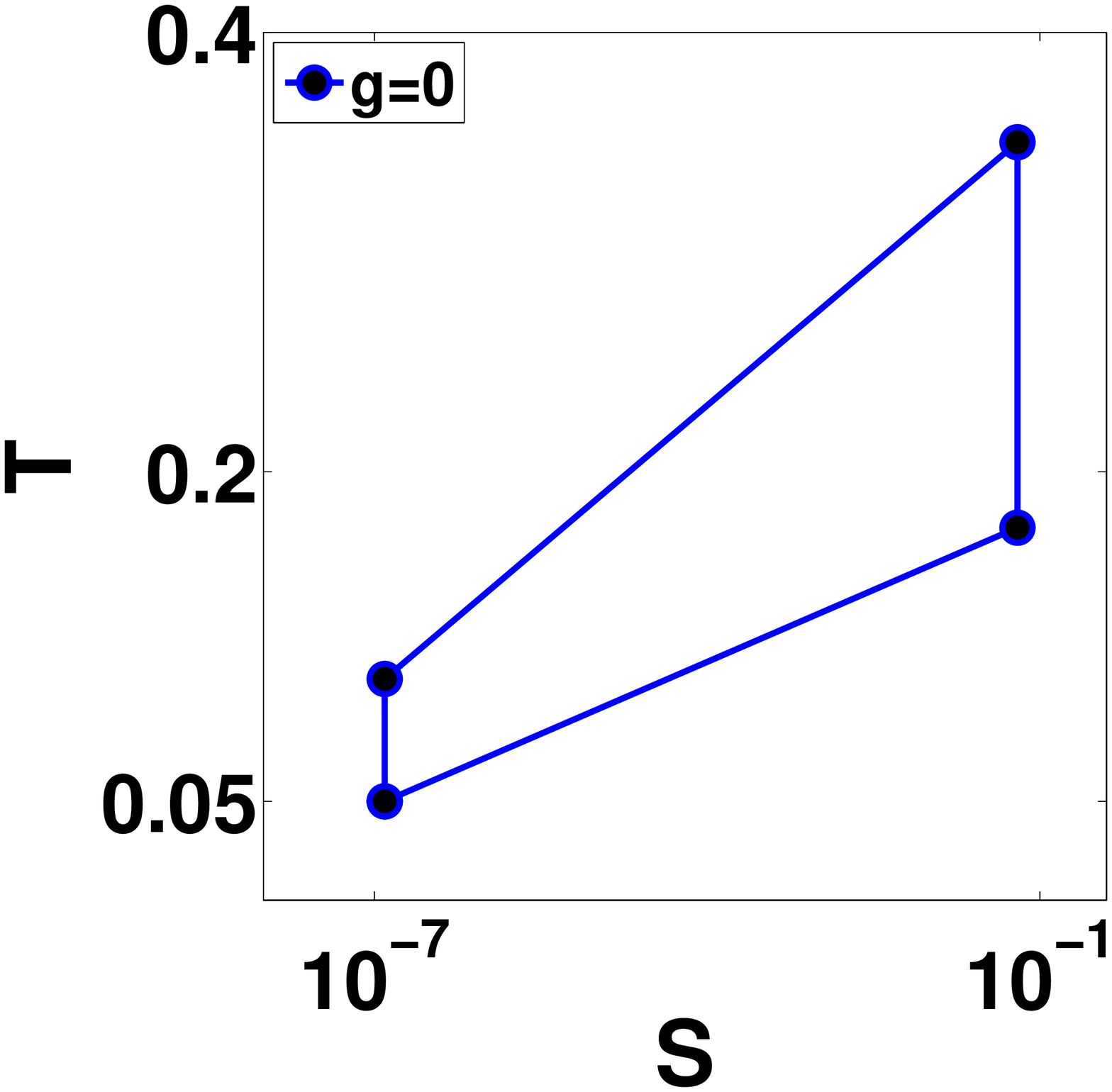}
        }%
    \end{center}
    \caption{\label{fig:cycles} (Color online) 
(a) $T-S$ diagram of the quantum Otto engine for the interacting case with $g=0.8$
and (b) for the non-interacting case with $g=0$.
The other parameters are taken as $\omega_h=2\omega$, $\omega_l=\omega$, $T_h=0.35$, $T_l=0.05$, 
and $\epsilon=0.005\omega$. 
$g$ and $T$ are in units of $\omega$ and $\hbar\omega/k$, respectively.}
\end{figure}

Classical and quantum Otto cycles have the same temperature and entropy ($T-S$) diagrams. We calculate the $T-S$ diagram
of the quantum Otto engine for interacting and non-interacting cases and compare them in Fig.~\ref{fig:cycles}. The area of the diagram
indicates the positive work done by the working substance and it is larger in the presence of interactions. We calculate the areas for the non-interacting and interacting cases as $W\sim 0.033$ and $W\sim 0.068$, respectively. The flatness of the upper and lower edges (isochoric stages) increase with the interctions as well. This makes the Otto cycle more closer to the Carnot cycle. Accordingly the efficiency also increased in the interacting case. Using classical thermodynamics expressions the Otto efficiencies for the interacting and non-interacting cases are determined to be $\eta\sim 0.75$ and $\eta\sim 0.52$, while the Carnot efficiency is $\eta\sim 0.85$. Such benefits of the interaction are not monotonic however. It is not trivial to answer how the work and efficiency depend on interaction, and how the quantum correlations and coherence established by the interaction influence the work harvesting capability of the working substance.
In the subsequent sections we explore the subtle effects of the interplay between the interactions, quantum coherence and quantum correlations on the magnitude and efficiency of work harvesting from classical resources.
\section{Quantum coherence and entanglement in generalized Rabi model}
\label{sec:qcoherence}
\begin{figure*}[t!]
     \begin{center}
        \subfigure[]{%
            \includegraphics[width=6cm]{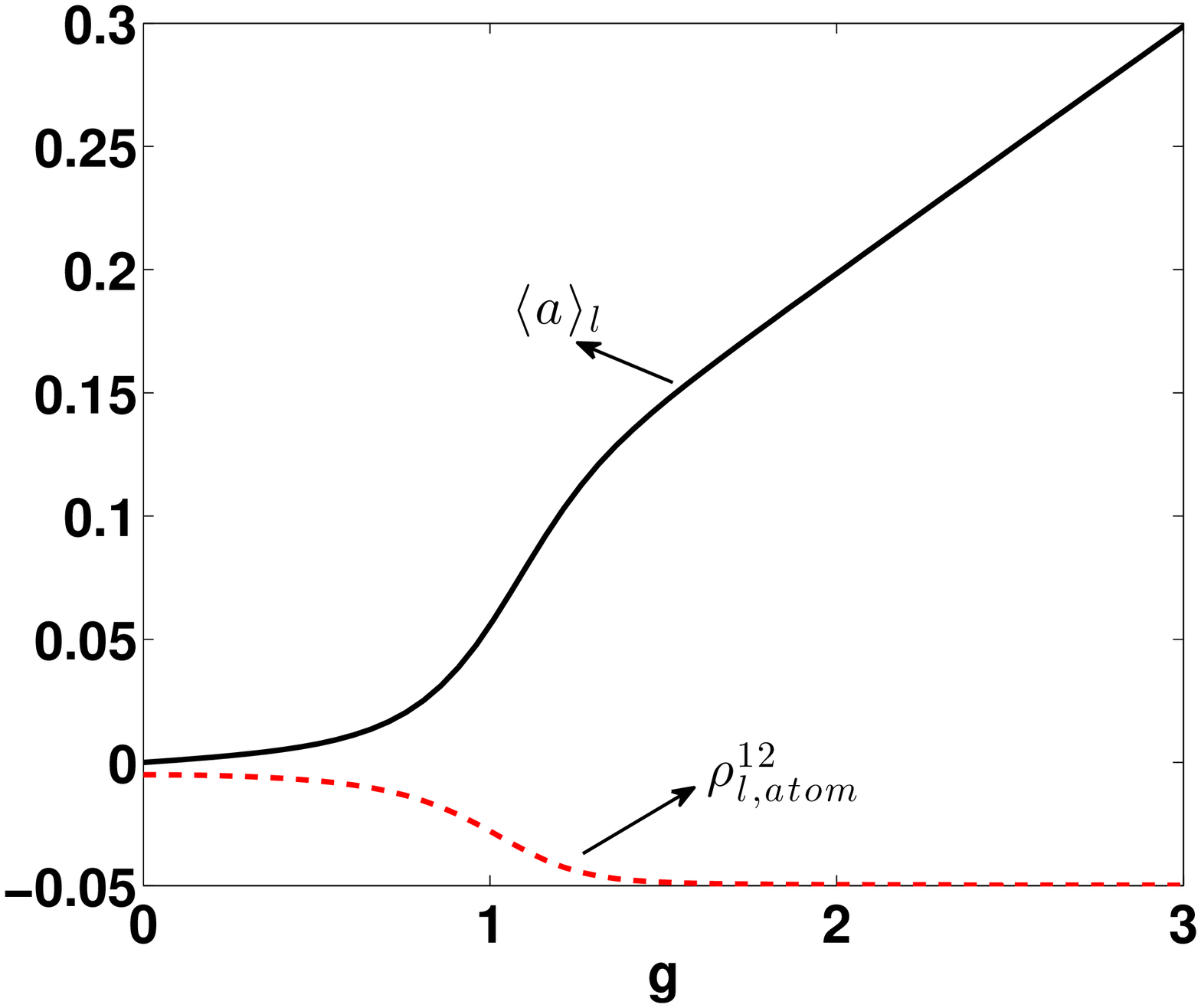}
        }%
        \subfigure[]{%
           \includegraphics[width=6cm]{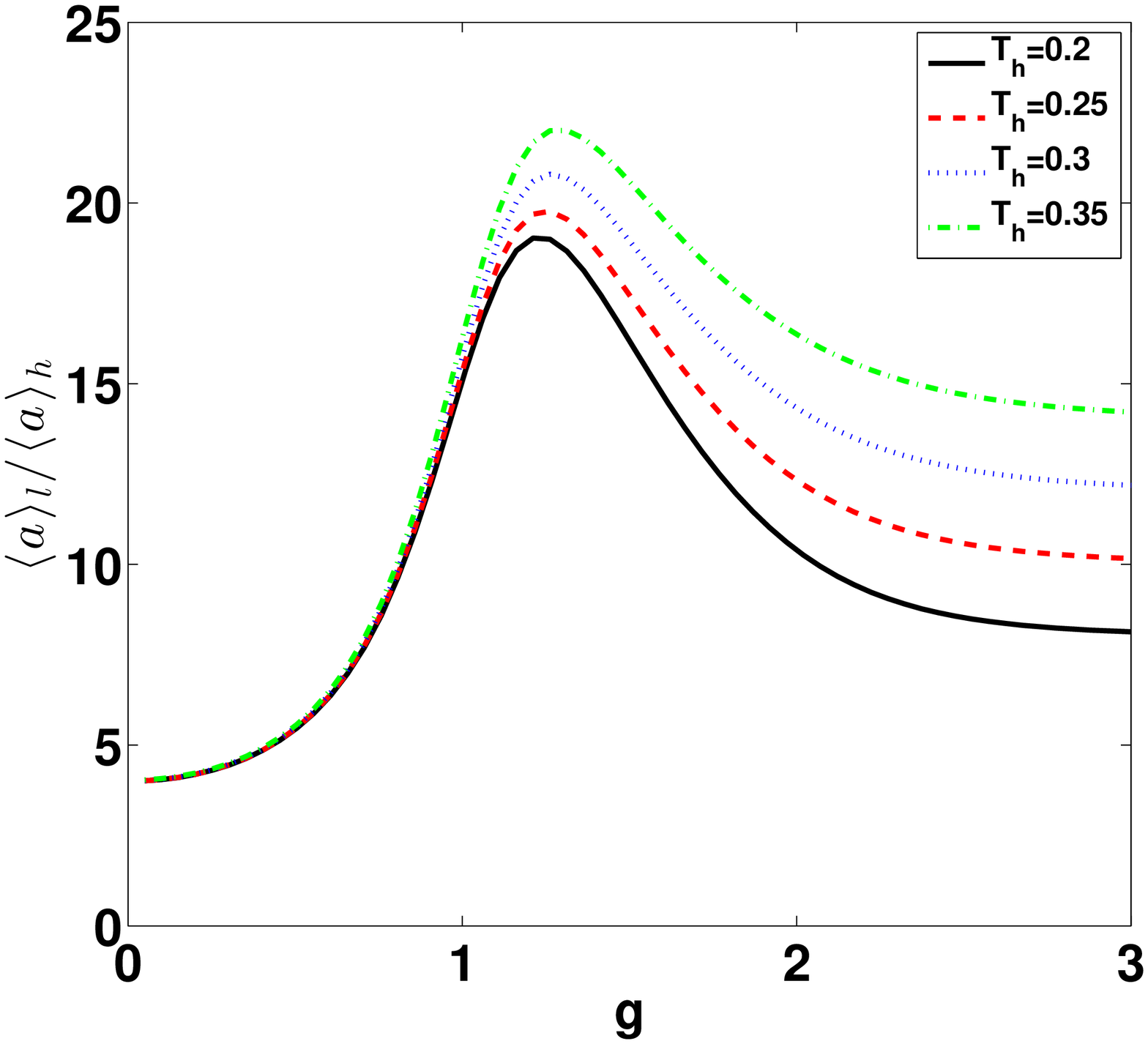}
        }%
	\subfigure[]{%
           \includegraphics[width=6cm]{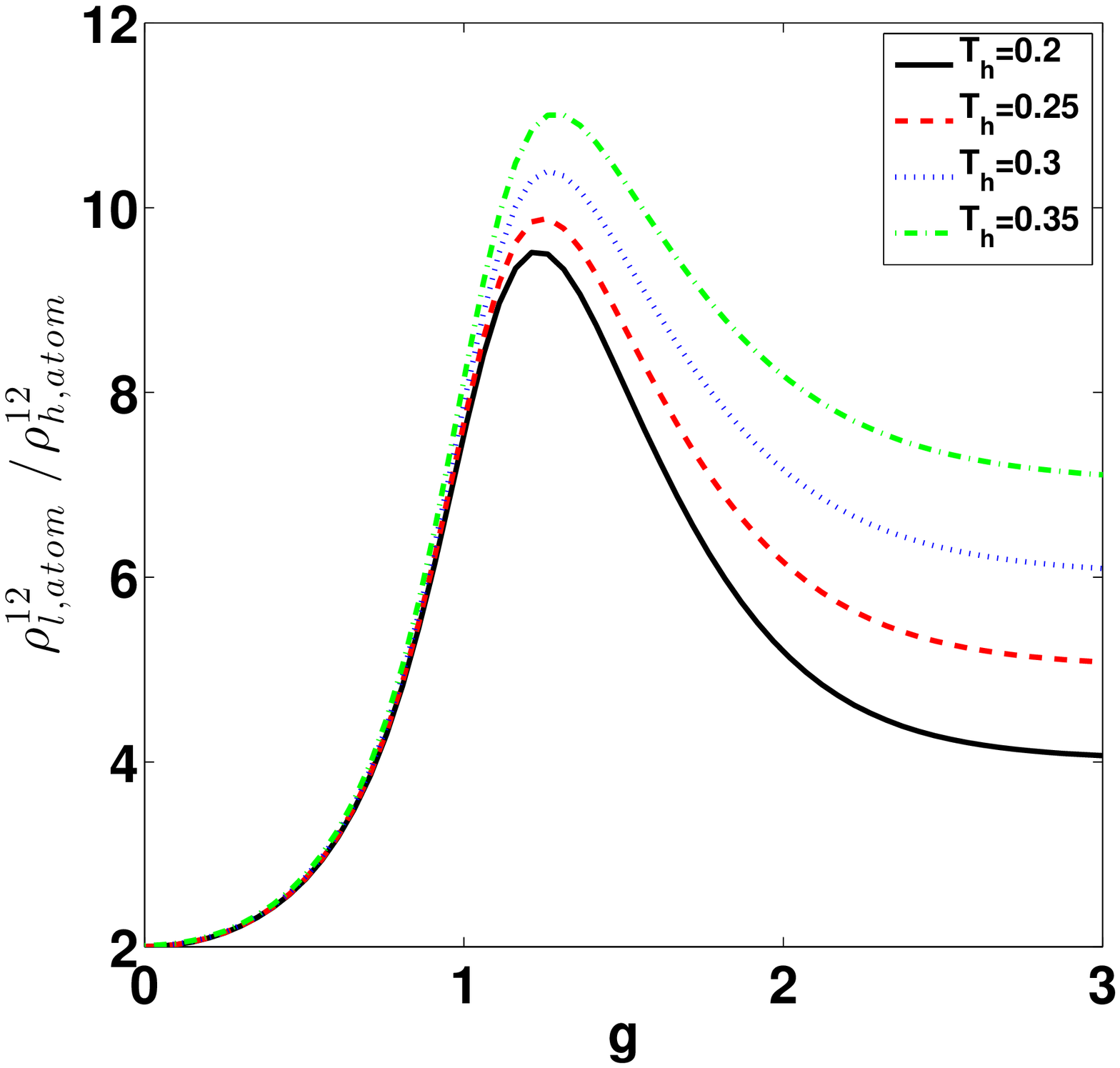}
        }%
	\caption{\label{fig:coherences} (Color online) 
	(a) Interaction constant $g$ dependence of quantum coherence in the central two-level system (atom) 
	$\rho_{l,\mathrm{atom}}^{12}$, red dashed curve) and 
	in the bosonic field ($\langle a \rangle_l$, black solid curve) at the end of the contact with the cold heat reservoir at $T_l=0.05$.  
	The relative quantum coherence (b) in the field $\langle a\rangle_l/\langle a\rangle_h$, and (c) in the two-level atom 
	$\rho_{l,\mathrm{atom}}/\rho_{l,\mathrm{atom}}^{12}$
	at the end of the contact with the cold heat reservoir at $T_l=0.05$ 
	with respect to the one at the end of the contact with the hot heat reservoir at temperatures 
	$T_h = 0.2$ (black solid curve), $0.25$ (red dashed curve), $0.3$ (blue dotted), $0.35$ (green dash dotted).
	The other parameters are taken as $\omega_h=2\omega$, $\omega_l=\omega$, and $\epsilon=0.005\omega$. 
	Coherence is dimensionless; $g$ and the temperatures $T_l,T_h$ are in units of $\omega$ and $\hbar\omega/k$, respectively.}
	\end{center}
\end{figure*}
The $Z_2$ symmetry breaking interaction in the generalized Rabi model induces a quantum superposition of the excited and 
ground states of the central two-level system (atom). This can be characterized by the examination of the off-diagonal terms of the 
reduced density matrix of the system, $\rho^{12}=Tr_{\mathrm{field}}(\rho)$, where trace is taken over the bosonic field degrees of freedom.
Bosonic field represent a weakly excited large two-level ensemble. Its coherence can be examined by calculating the expectation
value of the field annihilation operator $\langle a\rangle=\sum_n\langle \sigma_n\rangle/\sqrt{N}$.

We evaluate the  quantum coherences in the atom and in the field 
at the end of the contacts with the cold and hot heat reservoirs and report our results in Fig.~\ref{fig:coherences}.
The left panel shows that quantum coherence for the cold bath case
is monotonically increasing with the interactions 
both for the field ($\langle a\rangle_l$) and the central 
atom ($\rho^{12}_{l,\mathrm{atom}}$). When there is no interaction, there is only small seed coherence ($\epsilon$) in the atom. 
Without seeding coherence, the field cannot grow coherence with interactions. 

We find similar behavior for the hot bath case, but with smaller values than the ones for the cold bath case. The transition from
weak to large coherence also appears at higher threshold of interaction, around $g\sim 2\omega$.
Their ratios are shown in the middle and in the right panels. There is a critical interaction strength $g>
\sim\omega$ 
for which the relative
coherences are maximum. At higher hot reservoir temperatures $T_h$, relative coherence is higher 
for the same cold reservoir temperature $T_l$. Temperature dependence is weak for $g<\omega$. 
Sudden transition from weak to significant coherence in the field could be compared with the critical behavior
in the single atom superradiance~\cite{ashhab_superradiance_2013}. 

\begin{figure}[!t]
\begin{center}
\includegraphics[width=0.5\textwidth,angle=0]{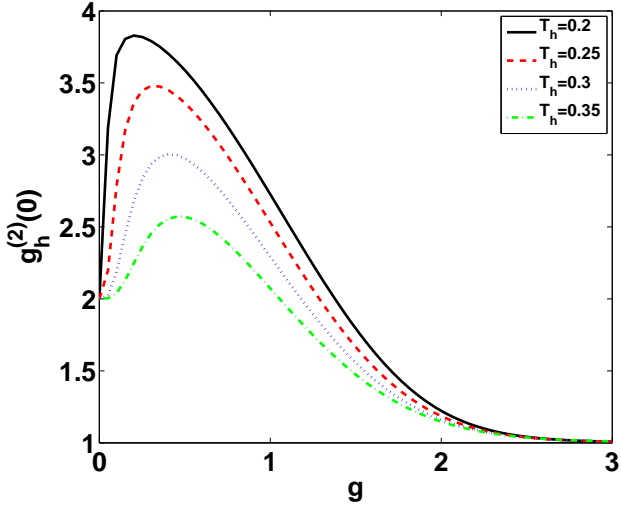}
\caption{\label{fig:g2} (Color online) 
Dependence of second order coherence function of the bosonic field $g^{(2)}_h(0)$ (dimensionless) on the interactions 
$g$ at the end of the contact with the hot thermal reservoir at  temperatures 
$T_h = 0.2$ (black solid curve), $0.25$ (red dashed curve), $0.3$ (blue dotted), $0.35$ (green dash dotted).
Second order coherence $g^{(2)}_l(0)$ for the cold bath case shows similar behavior but reaches to $1$ around $g>~\omega$. 
The other parameters are taken as $\omega_h=2\omega$, $\omega_l=\omega$, and $\epsilon=0.005\omega$. 
$g$ and the temperatures $T_h$ are in units of $\omega$ and $\hbar\omega/k$, respectively.}
\end{center}
\end{figure}

We have also examined the zero time delay 
second order coherence function of the field defined as $g^{(2)}(0)=\langle a^\dag a^\dag a a\rangle / 
\langle a^\dag a\rangle^2$~\cite{glauber_quantum_1963,loudon_non-classical_1980}. The result for the hot reservoir case is shown in Fig.~\ref{fig:g2}.
Second order coherence $g^2(0)_l$ for the cold bath stage shows similar behavior but reaches to $1$ around $g>\sim\omega$. 
This indicates that below $g\sim \omega$ field is in thermal coherent state both at the end of the hot and cold bath stages;
while it becomes quantum coherent state for both hot and cold bath cases when $g>\sim 2\omega$. In between
$g\sim \omega$ and $g\sim 2\omega$, the field changes statistical character from thermal coherent to profoundly quantum coherent at the
cold bath case while it remains in thermal coherent state for the hot bath case. We will see that these characteristic differences
associated with different operation regimes, namely the refrigerator (or heat pump) and heat engine regimes, of the Otto cycle. 

\begin{figure}[t!]
     \begin{center}
        \subfigure[]{%
            \includegraphics[width=7cm]{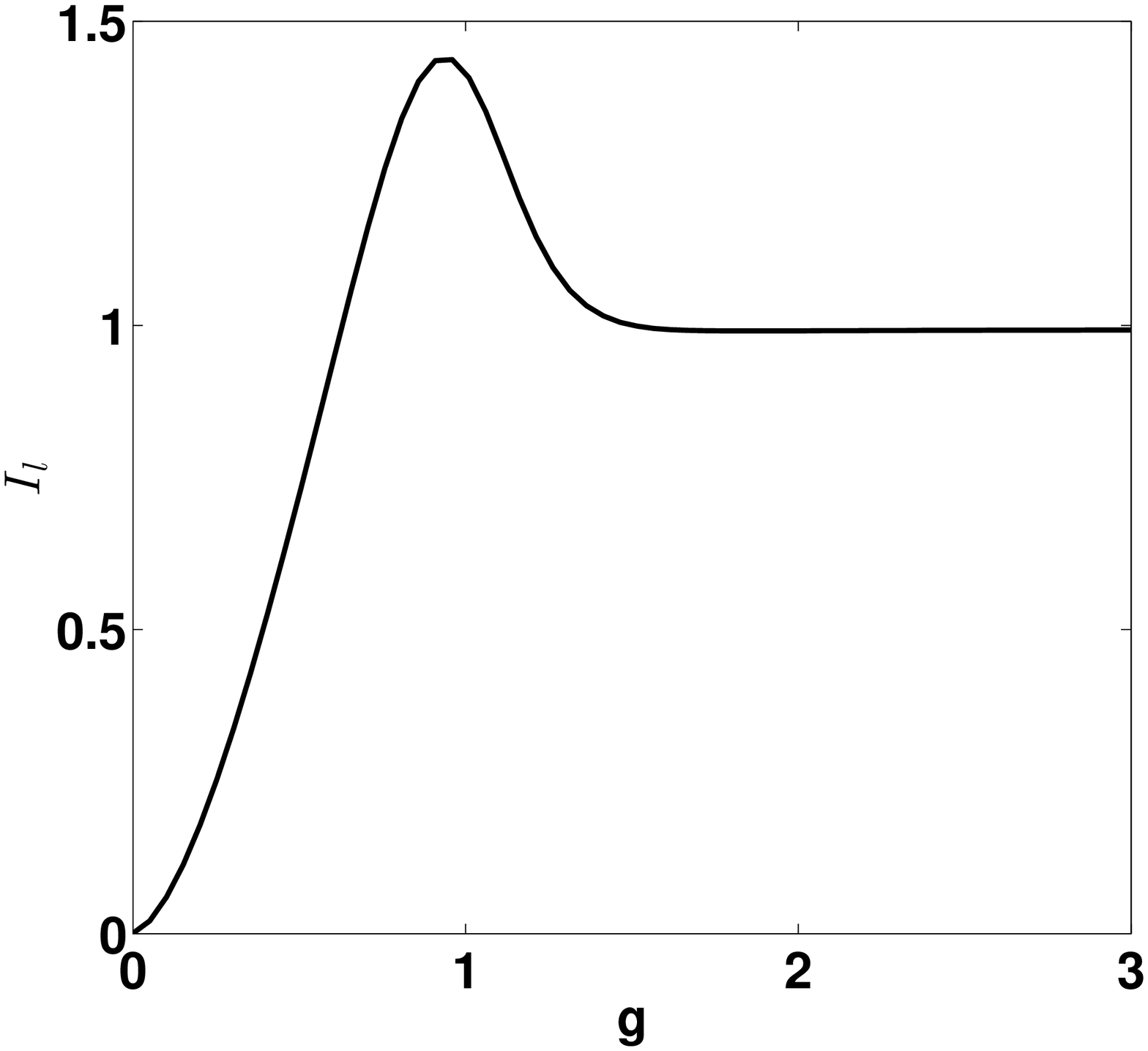}
        }\\%
        \subfigure[]{%
           \includegraphics[width=7cm]{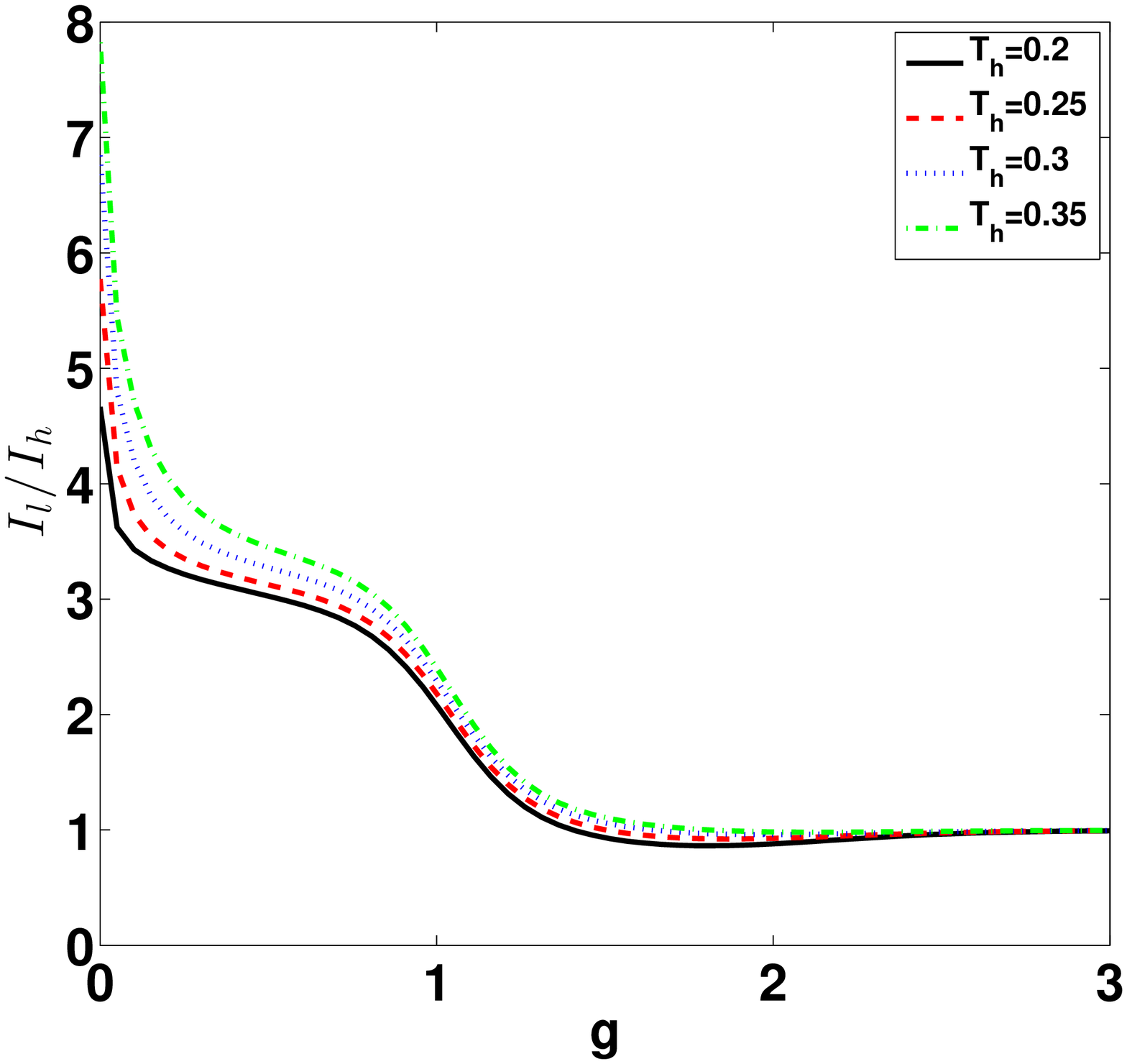}
        }%
	\caption{\label{fig:mutualInfo} (Color online) 
	(a) Interaction constant $g$ dependence of quantum mutual information $I_l$ at the end of the contact with the cold heat reservoir
	at temperature $T_l=0.05$.  (b) The relative quantum mutual information $I_l/I_h$ 
	at the end of the contact with the hot heat reservoir at temperatures 
	$T_h = 0.2$ (black solid curve), $0.25$ (red dashed curve), $0.3$ (blue dotted), $0.35$ (green dash dotted).
	The other parameters are taken as $\omega_h=2\omega$, $\omega_l=\omega$, and $\epsilon=0.005\omega$. 
	Coherence is dimensionless; $g$ and the temperatures $T_l,T_h$ are in units of $\omega$ and $\hbar\omega/k$, respectively.}
	\end{center}
\end{figure}

Long range quantum coherence in the ensemble of weakly excited two-level systems can establish quantum correlations in the
system. Total amount of quantum and classical correlations can be examined with the quantum mutual information~\cite{vedral_quantifying_1997} defined as 
$I=S(\rho_{\mathrm{atom}})+S(\rho_{\mathrm{field}})-S(\rho)$, where $S=-Tr(\rho\ln(\rho))$ is the 
von Neumann entropy of density matrix $\rho$. Atomic and field 
reduced density matrices are denoted by $\rho_{\mathrm{atom}}$ and $\rho_{\mathrm{field}}$, respectively.
We plot the behavior of quantum mutual information for the cold bath case $I_l$ and the relative quantum mutual information 
$I_l/I_h$  with the interactions $g$ in Fig.~\ref{fig:mutualInfo}. We see that in the $g<\sim\omega$ regime, there is strong relative
quantum mutual information in the system. In the $g>\sim2\omega$ and intermediate $\sim\omega<g<\sim2\omega$ regimes the quantum mutual 
information content in the
cold and hot bath cases become approximately the same. In the intermediate regime, hot bath case can have slightly larger quantum mutual information.

To reveal the content of this quantum mutual information in terms of quantum entanglement and other quantum correlations beyond entanglement, it is necessary to evaluate quantum correlation measures such as quantum discord~\cite{henderson_classical_2001,ollivier_quantum_2001} or entanglement of 
formation~\cite{wootters_entanglement_1998}. Such measures are not available for large and mixed quantum systems 
such as the one we consider here. Alternatively, we can investigate bipartite quantum entanglement between the central 
two-level site and the field representing the large ensemble by calculating the logarithmic negativity~\cite{vidal_computable_2002}. 
It is a non-convex entanglement monotone applicable to mixed states~\cite{plenio_logarithmic_2005} and defined as
$
E_N(\rho)=\log_2\vert\vert\rho^{\Gamma}\vert\vert_1,
$
where $\rho^{\Gamma}$ is the partial transpose with respect to a subsystem and $\vert\vert\rho^{\Gamma}\vert\vert_1$ is the trace norm of $\rho^{T}$.

\begin{figure}[t!]
     \begin{center}
        \subfigure[]{%
            \includegraphics[width=7cm]{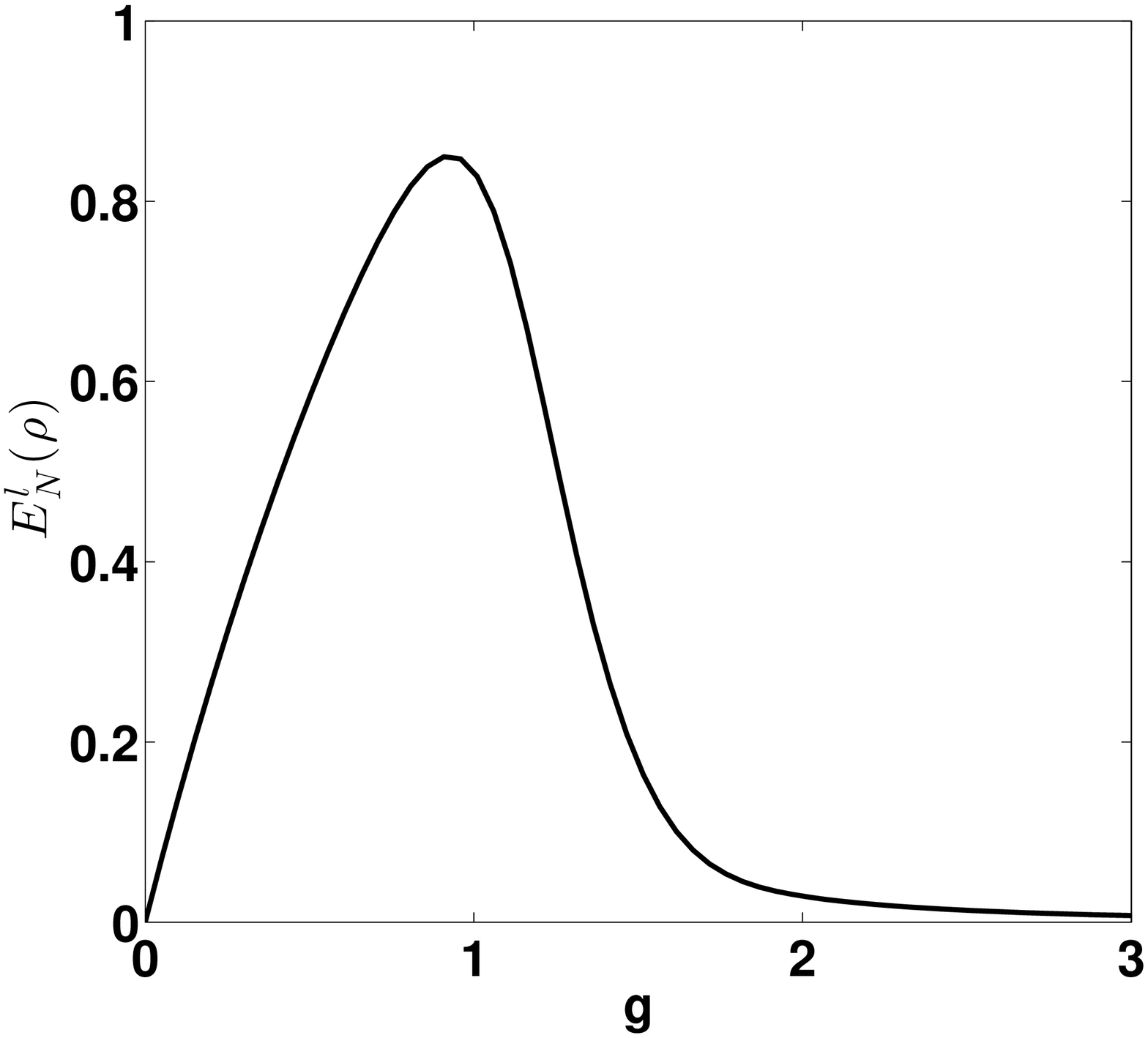}
        }\\%
        \subfigure[]{%
           \includegraphics[width=7cm]{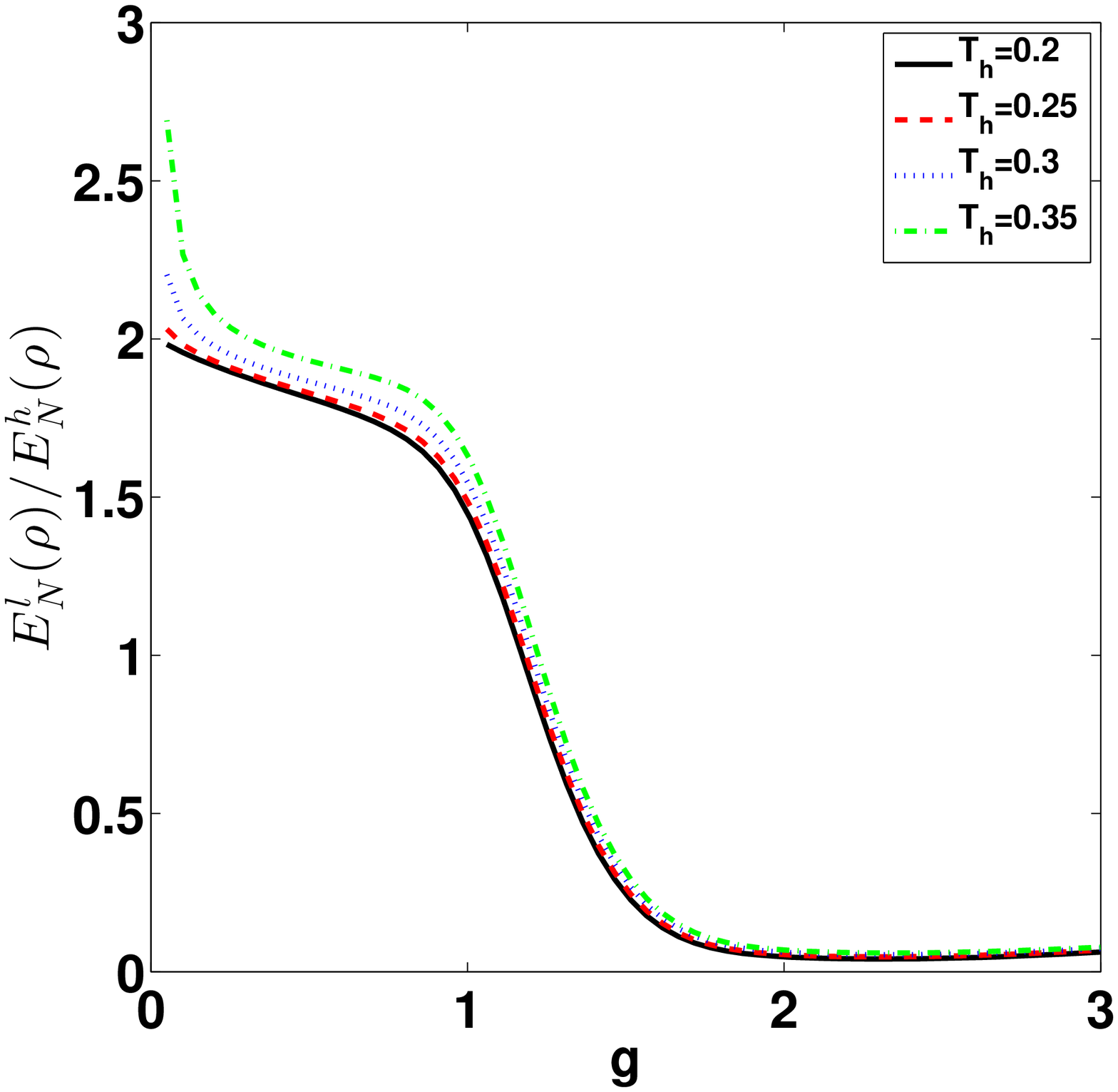}
        }%
	\caption{\label{fig:logNeg} (Color online) 
	(a) Interaction constant $g$ dependence of logarithmic negativity $E_N^l(\rho)$ at the end of the contact with the cold heat reservoir
	at temperature $T_l=0.05$.  (b)
	The relative logarithmic negativity $E_N^l(\rho)/E_N^h(\rho)$ 
	at the end of the contact with the hot heat reservoir at temperatures 
	$T_h = 0.2$ (black solid curve), $0.25$ (red dashed curve), $0.3$ (blue dotted), $0.35$ (green dash dotted).
	The other parameters are taken as $\omega_h=2\omega$, $\omega_l=\omega$, and $\epsilon=0.005\omega$. 
	Coherence is dimensionless; $g$ and the temperatures $T_l,T_h$ are in units of $\omega$ and $\hbar\omega/k$, respectively.}
	\end{center}
\end{figure}

Our results are given in Fig.~\ref{fig:logNeg}. We see that logarithmic negativity for the cold bath case is larger
than the one in hot bath case in $g<\omega$ regime. The ratio drops sharply in the intermediate regime 
$\sim\omega<g<\sim2\omega$ down
to zero and it remains zero in the $g>\sim 2\omega$  regime.

After analyzing the quantum coherence and entanglement in the working substance described by a generalized Rabi model
in a quantum Otto cycle, we will examine the work output and efficiency of the cycle in the preceeding section and determine
the regime of positive work and its relation to quantum coherence.
\section{Positive work regime and quantum coherence}

\begin{figure}[!t]
\begin{center}
\includegraphics[width=8cm,angle=0]{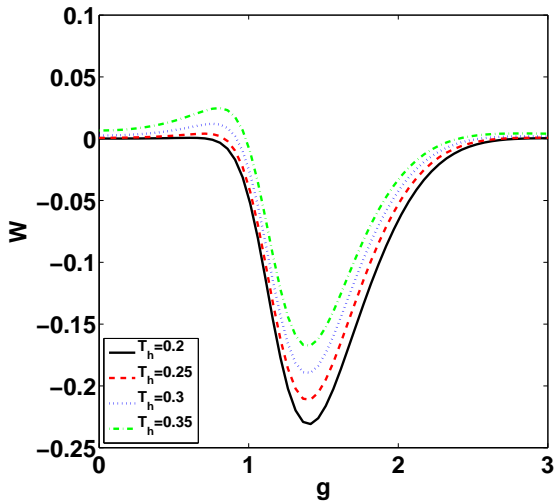}
\caption{\label{fig:workEff} (Color online) Work $W$ (in units of $\hbar\omega$) versus coupling strength $g$ (in units of $\omega$) for $\omega_h=2\omega$, $\omega_l=\omega$, $\epsilon=0.005\omega$, $T_l=0.05$, $T_h=0.2$ (black, solid line), $T_h=0.25$ (red, dashed line), $T_h=0.3$ (blue, dotted line) and $T_h=0.35$ (green, dot-dashed line). Here the temperatures are in units of $\hbar\omega/k$.}
\end{center}
\end{figure}

We have seen that seed coherence in the central site can be distributed over many weakly excited sites, represented by a bosonic
field in the continuum approximation, which leads to strong long range coherence. Towards the ultrastrong coupling regime,
interaction enhanced coherence builds strong quantum mutual information and bipartite entanglement with much higher amount
in the cold reservoir stage than the one in the hot reservoir case. In order to see if this can contribute to enhanced energy harvesting
and engine operation, we plot the net work output of the system in Fig.~\ref{fig:workEff}.
The work output is positive in the $g<\sim\omega$ and $g>\sim 2\omega$ regimes. Positivity, as well as the relative coherence, information and entanglement, grow with the thermal gradient over the entire interaction domain. 

Comparing the behavior of quantum mutual information in Fig.~\ref{fig:mutualInfo} and quantum entanglement in Fig.~\ref{fig:logNeg} with the positive work in Fig.~\ref{fig:workEff},
we see that heat engine operation can be obtained when there is strong quantum correlations at the end of the cold bath stage and when the system is more correlated in the cold reservoir stage than in the hot reservoir case. 
Despite the larger information gradient at low interaction coefficients in Figs.~\ref{fig:mutualInfo} and ~\ref{fig:logNeg}, the amount of quantum information is too weak for such regimes while around $g<\sim\omega$, both the quantum mutual information and bipartite entanglement for the cold reservoir stage reach their maximums. Just before the sudden drop of relative information and relative entanglement, the system could exploit large amount and large gradient of quantum correlations as a resource for positive work.
Our results generalize the recent studies in small coupled spin systems~\cite{huang_special_2014,zhang_entangled_2008,wang_thermal_2009} 
where local maximum work is reported at a critical interaction strength, relative coherence 
or at a critical amount of entanglement; and illuminate the underlying role of simultaneously optimized relative entanglement and
amount of entanglement in the cold bath case. In addition, the interplay of quantum coherence and quantum mutual information to
build up quantum information resources for enhanced engine operation is revealed. 

Our conclusion, which is based upon on observation of the figures, provides additional evidences on the possible relation between work and efficiency and quantum correlations and coherences. An explicit relation of the work produced by the engine in terms of quantum correlations shows that the work output is proportional to the difference of the amount of quantum entanglement, characterized by the concurrence, between the cold and hot stages of the engine cycle~\cite{zhang_entangled_2008,wang_thermal_2009,zhang_four_2007}. Our Rabi model is based upon pairwise interaction of a single spin with spins from a large ensemble. Our conclusion is an intuitive generalization of the two spin result to our case of large spin ensemble. Instead of concurrence gradient, our system benefits from gradient of many body quantum correlations, characterized by mutual information or logarithmic negativity. 

We could give a more intuitive explanation. The larger the gradient of quantum correlations, the larger amount of heat is absorbed from the hot reservoir to reduce the quantum correlations. Large quantum correlations gradient then contribute positively to the harvested work in the expansion stage. Larger correlations, in addition to large gradient, make their contribution more significant. The interplay of thermal gradient and correlation gradient would be within the boundaries of the second law and the positive work condition would be violated at a critical gradient beyond which the engine would turn into a refrigerator or a  heat pump. This is analytically shown for the case of two spins~\cite{zhang_entangled_2008,wang_thermal_2009,zhang_four_2007} and our spin ensemble results are in full accordance with the two spin theory.

After a sharp drop around $g<\sim\omega$, total correlations in the cold and hot reservoir stages become approximately same while the quantum entanglement in the hot reservoir case becomes relatively larger. In that case, entanglement in the hot bath case contributes as a resource for negative work. Maximum negative work is at a critical point where the amount of entanglement and its relative strength are both large in the hot bath case. As temperature of the hot bath increases, amount of entanglement decreases and work becomes less negative. Role of entanglement on the performance in smallest possible self-contained quantum fridge has been discussed recently and it has been found that such smallest fridges can cool to lower temperatures in the presence of entanglement~\cite{brunner_entanglement_2014}; though in some cases entanglement could not be used as a resource~\cite{correa_performance_2013}. Our analysis points out the significance of simultaneous optimization of 
relative entanglement and amount of entanglement in a larger system to operate between refrigerator or heat pump 
and heat engine phases. In addition the interplay of quantum coherence and von Neumann mutual information to build
quantum information resource for such systems are revealed.

On the other hand, it is difficult to reach such high interaction regimes in available practical systems, where negative work could be used for refrigeration or heat pump applications. The heat engine phase in the 
interaction regimes up to ultrastrong coupling, $g<\sim\omega$, and the associated 
generalized Rabi model however can already be realized in modern spin-boson systems such as in circuit QED. In the following we consider quantum heat engine operation and analyse work and efficiency with more detail.
\section{Efficiency and positive work of quantum coherent Otto heat engine}
\label{sec:results}
\subsection{Changing resonance frequency in adiabatic stages}
\begin{figure}[!t]
\begin{center}
\includegraphics[width=8.0cm]{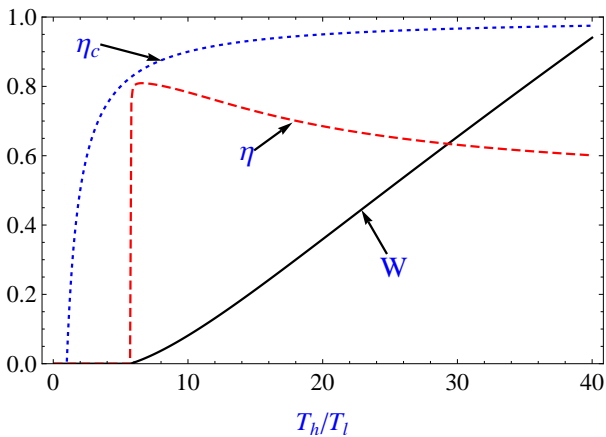}
\caption{\label{fig1} Work $W$ (in units of $\hbar\omega$) and efficiency $\eta$ as functions of heat bath temperature $T_h$ for $T_l=0.05$ (in units of $\hbar\omega/k$) for $\omega_h=2\omega$, $\omega_l=\omega$, $\epsilon=0.005\omega$ and $g=0.9\omega$. The uncoupled working medium ($g=0$) has an efficiency $\eta\approx 0.5$ for $\omega_h/\omega_l=2$.}
\end{center}
\end{figure}

We analyze our quantum Otto engine in the ultrastrong coupling regime and focus on the engine performance by changing the resonance frequency in the adiabatic branches of the cycle at a fixed coupling strength, $g_h=g_l=g$. In Fig.~\ref{fig1} we plot the net work output $W$ and the efficiency $\eta$ of the engine as functions of hot reservoir temperature $T_h$ for cold reservoir temperature $T_l=0.05$ (in units of $\hbar\omega/k$), respectively for $g/\omega=0.9$, $\omega_h=2\omega$ and $\omega_l=\omega$. Similar qualitative behavior is found for other $T_l$. In the case of uncoupled subsystems ($g=0$), the net work is equal to the sum of the work performed by each subsystems; neglecting small $\epsilon$, the efficiency and the positive work condition are determined by $\eta\approx 1-\omega_l/\omega_h$ and $T_h>qT_l$ with $q\approx \omega_h/\omega_l$, respectively. First order corrections in $\epsilon$ to the approximate expressions are in the order of $10^{-5}$ for $\epsilon=0.005\omega$. In the case of coupled subsystems the positive work condition for the engine requires larger temperature ratios than the case of uncoupled engine as can be seen in Fig.~\ref{fig1}. The work output is high at high temperature regions by the contribution of higher energy levels. The efficiency of the coupled engine is larger $(\eta_{\max}\approx 0.8)$ than the uncoupled one only at the low temperatures. 

\begin{figure}[!t]
\begin{center}
\includegraphics[width=0.45\textwidth,angle=0]{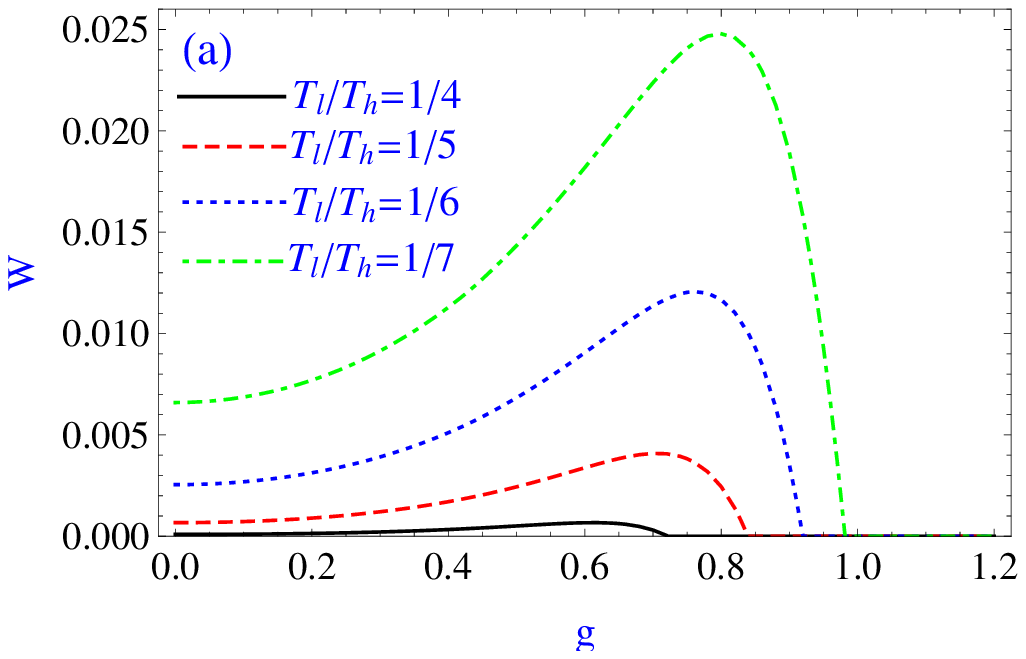}
\includegraphics[width=0.45\textwidth,angle=0]{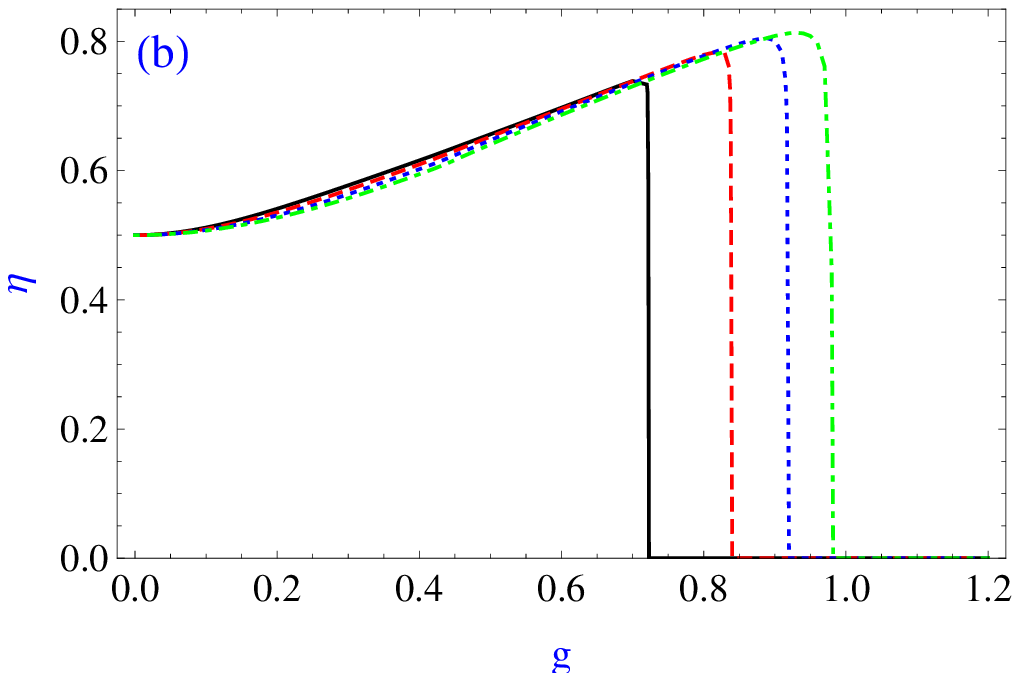}
\caption{\label{fig2} (Color online) Work $W$ (a) (in units of $\hbar\omega$) and efficiency $\eta$ (b) versus coupling strength $g$ (in units of $\omega$) for $\omega_h=2\omega$, $\omega_l=\omega$, $\epsilon=0.005\omega$, $T_l=0.05$, $T_h=0.2$ (black, solid line), $T_h=0.25$ (red, dashed line), $T_h=0.3$ (blue, dotted line) and $T_h=0.35$ (green, dot-dashed line). Here the temperatures are in units of $\hbar\omega/k$.}
\end{center}
\end{figure}

Further analysis of temperature and coupling strength effects on the performance of the QHE has been given in Fig.~\ref{fig2} for  low temperatures, where we have plotted the net work done by the engine and its efficiency as functions of coupling strength. In the Jaynes-Cummings regime where $g/\omega<<1$, the coupling strength has no significant contribution to the engine performance than the uncoupled one. On the other hand, if the coupling regime is changed from weak to ultrastrong ($g\sim\omega$) regime, the engine efficiency can be notably improved. Both the net work output and efficiency can increase with the increase in $g/\omega$. It is remarkable that for the regions where work increases with $g/\omega$, $W$ is an increasing function of $\eta$. The ratio $T_h/T_l$, which determines the Carnot efficiency ($\eta_c=1-T_l/T_h$), has also a decisive role in the performance of our QHE. The higher ratio in $T_h/T_l$ suggests higher net work output and maximum efficiency, and at the same time allows for the engine to operate in a more wide range of coupling constant. In particular, for the ratios $T_l/T_h=0.25$ and $T_l/T_h=0.2$ the engine efficiency increases with $g$ close to Carnot bound beyond which it sharply drops to zero and coupled engine performs less efficient than the uncoupled one. On the other hand, the engine operation is changed from heat engine to either refrigerator or heat pump and again relative to uncoupled refrigerators or heat pumps their cooling or warming efficiencies would be larger. Similar qualitative behavior for the efficiency is found for simpler interacting spin engines~\cite{thomas_coupled_2011}. 
\subsection{Changing interaction constant in adiabatic stages}
\begin{figure}[!t]
\begin{center}
\includegraphics[width=8.0cm]{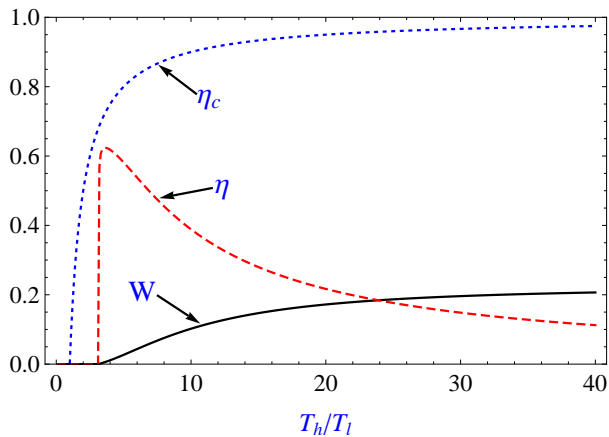}
\caption{\label{fig3} Work $W$ (in units of $\hbar\omega$) and efficiency $\eta$ as functions of heat bath temperature $T_h$ for $T_l=0.05$ (in units of $\hbar\omega/k$) for $\omega_h=\omega_l=\omega$, $\epsilon=0.005\omega$, $g_h=0.4\omega$ and $g_l=0.9\omega$.}
\end{center}
\end{figure}

Now we consider an alternative adiabatic processes where the coupling strength is changed between two values $(g_h\rightarrow g_l\rightarrow g_h)$ for a fixed resonance frequency $\omega_h=\omega_l=\omega$. Fig.~\ref{fig3} shows the thermodynamic quantities as functions of heat bath temperature $T_h$ for $T_l=0.05$ (in units of $\hbar\omega/k$)  for $g_h=0.4\omega$ and $g_l=0.9\omega$. $W$ and $\eta$ have nearly the same qualitative behavior as in Fig.~1. For the temperature regimes where the quantum nature of the working medium is pronounced, the efficiency is high, while for the temperatures where high energy levels become populated the work output is significant though less than the case of changing frequency. 

\begin{figure}[!t]
\begin{center}
\includegraphics[width=0.45\textwidth,angle=0]{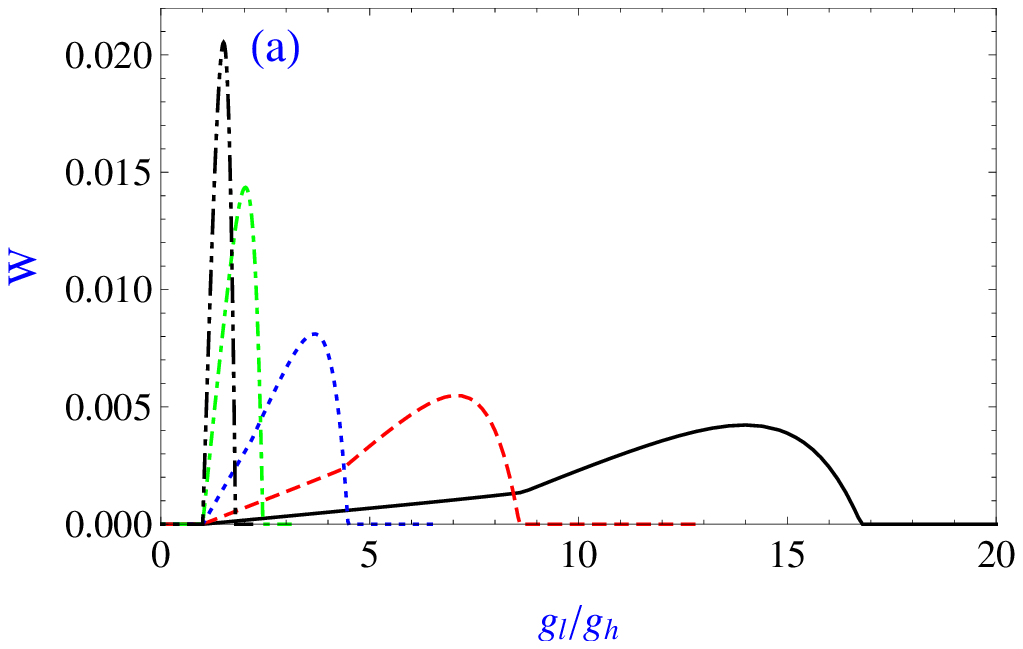}
\includegraphics[width=0.45\textwidth,angle=0]{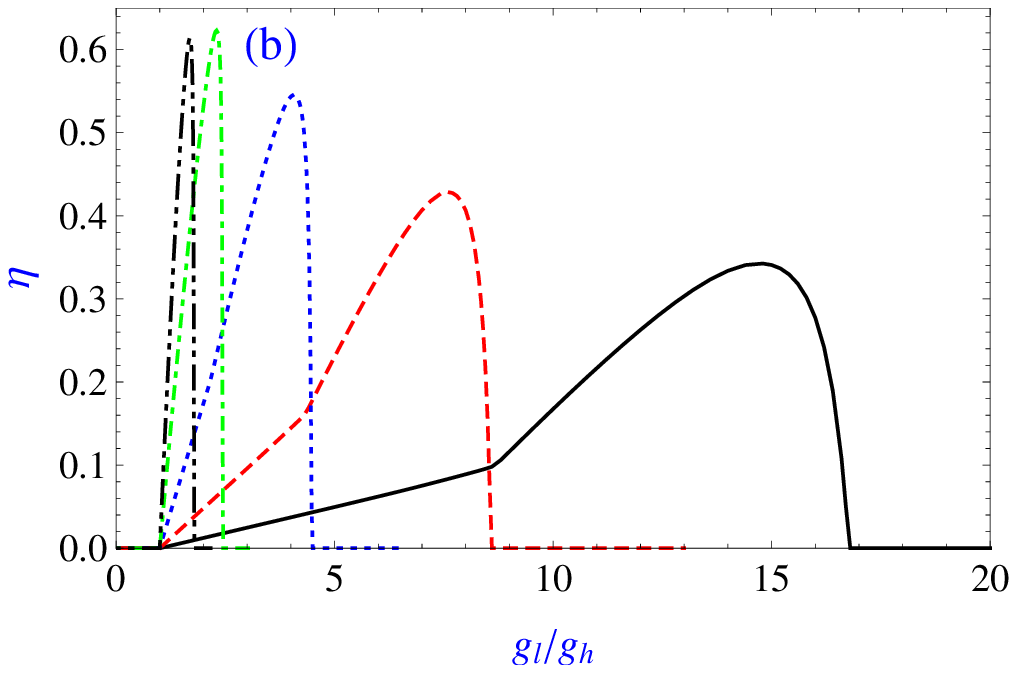}
\caption{\label{fig4} (Color online) Work (a) (in units of $\hbar\omega$) and efficiency (b) versus $g_l$ (in units of $\omega$) for $\omega_h=\omega_l=\omega$, $T_h=0.2$, $T_l=0.05$ (in units of $\hbar\omega/k$), $\epsilon=0.005\omega$, $g_h=0.05\omega$ (black, line), $g_h=0.1\omega$ (red, dashed line), $g_h=0.2\omega$ (blue, dotted line), $g_h=0.4\omega$ (green, dot-dashed line) and $g_h=0.6\omega$ (black, dot-dot-dashed line).}
\end{center}
\end{figure}

In Fig.~\ref{fig4}, we have shown a more detailed investigation for the role of coupling changes on the coupled QHE at the low temperature case with $T_l/T_h=0.25$ (where $\eta_c=0.75$). The engine can produce work with high efficiency (up to $\eta\approx 0.63$) when the changes in the coupling strength are done in the ultrastrong level. On the other hand, when the changes in the adiabatic branches are at the Jaynes-Cummings (weak coupling) regime, the engine performance is useless. Specifically when $g_l,g_h>0.4\omega$ the performance of the QHE can be dramatically increased. Energy spectrum of the Rabi model in Ref.~\cite{ridolfo_thermal_2013} shows that the lowest level is almost flat for small coupling coefficients. It drops sharply after $g/\omega\approx 0.42$. Such a steep variation of the lowest energy level with $g$ in the ultrastrong coupling regime is beneficial for positive work gradient with the change $g_l>g_h$ after this critical value at low temperatures. Both $W$ and $\eta$ have a single maximum slightly at different $g_l$ values. Therefore, except a small $g_l$ range, the net work output increases with the efficiency. Fig.~\ref{fig4} also demonstrates that the engine cannot produce work when $g_l<g_h$, as the first energy gap which is dominant in the work extraction, becomes subject to quantum adiabatic contraction in the stage 2 of the 
cycle by the change $g_l<g_h$, instead of expansion. 

\section{Possible Implementation with circuit QED}
\label{sec:Implementation}

Before we conclude, we remark that the generalized Rabi model is experimentally available in various modern systems such as in optomechanics~\cite{sornborger_superconducting_2004} or in circuit QED~\cite{niemczyk_circuit_2010} (For other physical and experimental systems see e.g. Ref.~\cite{braak_integrability_2011} and references therein).

Here we consider a possible circuit QED implementation to estimate a few figure of merits for our proposed quantum coherent 
Otto heat engine by using readily available experimental parameters of the ultrastrong coupling regime of circuit QED~\cite{niemczyk_circuit_2010,forn-diaz_observation_2010}. Central two-level site can be implemented by a flux qubit and the representative bosonic field of the weakly excited ensemble of two-level sites can be effectively implemented by a 
superconducting waveguide resonator mode. Tunable ultrastrong coupling has been proposed for a multiloop qubit system~\cite{peropadre_switchable_2010}. Tunable coupling between a flux qubit and a microwave resonator has been demonstrated experimentally~\cite{gustavsson_driven_2012}. 
The magnetic energy of the qubit and the qubit gap correspond to $\epsilon$ and $\Delta$, respectively. For a typical resonator frequency  $\omega/2\pi \sim 6-9$ GHz we find $\epsilon\sim 0.005\omega\sim 30-45$ MHz, which is around the edge of the qubit linewidth~\cite{forn-diaz_observation_2010}. The qubit bare energy $\omega$ can be tuned over a wide range by an external magnetic flux. The coupling coefficient 
could be varied in the range $g/2\pi\sim 10$ MHz \cite{wallraff_strong_2004} - $1$ GHz~\cite{forn-diaz_observation_2010}; while theoretical studies suggest that $g$ can be up to $g\sim 3\omega$~\cite{peropadre_nonequilibrium_2013}. 

Thermal baths can be implemented by using a dilution refrigerator to cool the resonator to empty resonator limit and applying a thermal radiation at the input of the transmission line resonator which can lead to temperatures in the range of $\sim 100 - 7500$ mK~\cite{fink_quantum--classical_2010}. For the optimal low T for our QHE we find $T_l<1\sim 300 - 450$ mK. 

 In order to estimate the power $P$ produced by the engine, we can assume the adiabatic stages take time $\tau_{2,4}\gg 1/E$ where $E$ is a typical energy scale for the system, which can be taken here as $\sim$ GHz. 
Thermalization rate is in the order of MHz as the resonator decay rate is about $\kappa\sim 2\pi\times 1$ MHz. Accordingly, the cycle time is determined predominantly by the adiabatic stages~\cite{liao_single-particle_2010}.
Typical values of work output at $\eta\approx 0.6$ are $W\sim 0.02\omega\sim 10^{-25}$ J in changing $g$ scheme and $W\sim 0.6\omega\sim 10^{-24}$ J in changing $\omega$ scheme (for $\omega/2\pi = 9$ GHz), respectively so that they yield $P\ll 10^{-16} - 10^{-15}$ J/s, which is potentially promising a higher performance than those of single-ion Otto engines~\cite{abah_single-ion_2012}. Optimization of heat bath noise models and adiabatic processes via detailed finite time analysis can lead to higher powers and
efficiency can beat the Carnot limit~\cite{huang_effects_2012,rosnagel_nanoscale_2014,correa_quantum-enhanced_2014}. 
These considerations suggest that modern circuit QED systems can be used to implement quantum coherent Otto engine 
we proposed here.
\section{Conclusion}
\label{conc}
In conclusion, we proposed a quantum coherent Otto engine which can build strong quantum coherence and correlations and utilize them for enhanced positive work output efficiently. Our proposal is based upon a biomimetic strategy to simplify long range quantum coherence production 
mechanism of a light harvesting complex for synthetic systems. Instead of coupling all the pigments, we consider a central two-level 
site with a weak seed coherence coupled to the surrounding two-level sites. In the limit of weak excitation and many surrounding sites,
the system can be described by a generic spin-boson model known as generalized Rabi model. Central coherence is shared and amplified by coupling to many two-level sites as a quantum analog of classical synchronization of Huygen's clocks. Our calculations indicate that
such a working substance described by a generalized Rabi model, can operate close to Carnot efficiency in a quantum Otto cycle 
at low temperatures in ultrastrong coupling regime and produce work enhanced by interaction amplified quantum coherence and correlations.

We determine the positive work regime for heat engine operation as well as negative work regime for refrigerator or heat pump operations.
The interplay of quantum coherence and quantum correlations to determine these regions and to enhance work and efficiency is revealed
by comparing the interaction dependence of the quantum coherence, second order coherence, quantum mutual information and logarithmic negativity with that of work and efficiency.  
In general, the amount of quantum correlations at the end of the cold reservoir stage should be large, and relatively higher than those at the end of the hot reservoir stage. Our conclusions and proposed Rabi model are fundamental and applicable to various spin-boson systems as well as to large spin ensembles. As a concrete and currently feasible physical system, we proposed that a tunable superconducting circuit QED system can be used to implement our ideas. Our estimations of the operation time and work output lead to competitive figures of merit in the ultrastrong coupling regime of the tunable circuit QED system of a flux qubit and a superconducting transmission line resonator. 
%
\begin{acknowledgements}
The authors warmly thank N.~Allen for fruitful discussions. F.~A.~acknowledges the support and hospitality of the Off{\.i}ce of Vice President for Research and Development (VPRD) and Department of Physics of the Ko\c{c} University. A.~\"U.~C.~H. acknowledges the COST Action MP1209. \"O.~E.~M.~ and A.~\"U.~C.~H.~acknowledge the support by Lockheed Martin Corporation Grant.
\end{acknowledgements}

\end{document}